\documentclass[12pt]{article}

\usepackage{times}
\usepackage{fixltx2e}
\usepackage{graphicx}
\usepackage[table]{xcolor}
\usepackage{ragged2e}
\usepackage{multirow}
\usepackage{helvet}
\usepackage{booktabs}
\usepackage{colortbl}
\usepackage{amsmath}
\usepackage{cmbright}
\usepackage{hyperref}
\hypersetup{
    colorlinks=true,
    linkcolor=magenta,
    filecolor=magenta, 
    citecolor=sared,
    urlcolor=red,
}

\usepackage{blindtext}

\makeatletter
\newenvironment{figurehere}
{\def\@captype{figure}}
{}
\makeatletter

\usepackage{ccaption}
\usepackage{siunitx}
\usepackage{fixltx2e}
\newenvironment{sciabstract}{%
\begin{quote} \bf}
{\end{quote}}

\definecolor{sared}{rgb}{0.69, 0,0}

\newcounter{lastnote}

\title{\textbf{A Remedy to Compute-in-Memory with \\ Dynamic Random Access Memory:\\ 1FeFET-1C Technology for Neuro-Symbolic AI}}

\author
{Xunzhao Yin$^{1}$, Hamza Errahmouni Barkam$^{2}$, Franz M{\"u}ller$^{3}$, Yuxiao Jiang$^{1}$, \\
Mohsen Imani$^{2}$,
Sukhrob Abdulazhanov$^{3}$, Alptekin Vardar$^{3}$,  Nellie Laleni$^{3}$, \\
Zijian Zhao$^{4}$, Jiahui Duan$^{4}$, Zhiguo Shi$^{1}$,  
Siddharth Joshi$^{4}$,
Michael Niemier$^{4}$,  \\
X. Sharon Hu$^{4}$, Cheng Zhuo$^{1*}$, Thomas K{\"a}mpfe$^{3*}$, Kai Ni$^{4*}$
\\
\normalsize{$^{1}$Zhejiang University, Hangzhou, China;}\\
\normalsize{$^{2}$University of California Irvine, Irvine, USA;}\\
\normalsize{$^{3}$Fraunhofer IPMS, Dresden, Germany;}\\
\normalsize{$^{4}$University of Notre Dame, Notre Dame, USA;}
\\
\normalsize{$^\ast$To whom correspondence should be addressed; E-mail:}\\ \normalsize{czhuo@zju.edu.cn, thomas.kaempfe@ipms.fraunhofer.de, kni@nd.edu.}
}

\date{}

\begin{document} 

\maketitle 
\begin{sciabstract}
Neuro-symbolic artificial intelligence (AI) excels at learning from noisy and generalized patterns, conducting logical inferences, and providing interpretable reasoning. Comprising a "neuro" component for feature extraction and a "symbolic" component for decision-making, neuro-symbolic AI has yet to fully benefit from efficient hardware accelerators. Additionally, current hardware struggles to accommodate applications requiring dynamic resource allocation between these two components. To address these challenges—and mitigate the typical data-transfer bottleneck of classical Von Neumann architectures—we propose a ferroelectric charge-domain compute-in-memory (CiM) array as the foundational processing element for neuro-symbolic AI. This array seamlessly handles both the critical multiply-accumulate (MAC) operations of the "neuro" workload and the parallel associative search operations of the "symbolic" workload. To enable this approach, we introduce an innovative 1FeFET-1C cell, combining a ferroelectric field-effect transistor (FeFET) with a capacitor. This design, overcomes the destructive sensing limitations of DRAM in CiM applications, while capable of capitalizing decades of DRAM expertise with a similar cell structure as DRAM, achieves high immunity against FeFET variation—crucial for neuro-symbolic AI—and demonstrates superior energy efficiency. The functionalities of our design have been successfully validated through SPICE simulations and prototype fabrication and testing. Our hardware platform has been benchmarked in executing typical neuro-symbolic AI reasoning tasks, showing over 2x improvement in latency and 1000x improvement in energy efficiency compared to GPU-based implementations.

\end{sciabstract}

\section*{\textcolor{sared}{Introduction}}
\label{sec:introduction}

Neuro-symbolic artificial intelligence (AI) signifies an integrated approach to AI, fusing two divergent paradigms: the rule-based symbolic reasoning and the pattern-detecting neural networks \cite{hitzler2022neuro,sarker2021neuro,garcez2023neurosymbolic}. 
The evolving AI landscape acknowledges the insufficiencies of relying exclusively on either method. 
While powerful deep learning models frequently encounter challenges in scenarios necessitating reasoning, transparency, and sparse-data handling, conversely, symbolic AI tends to falter amidst noisy data and pattern generalization tasks. Neuro-symbolic AI synthesizes the merits of both approaches, striving to build systems that learn from environmental cues, draw logical inferences, and offer intelligible decision explanations.
This not only augments the robustness and transparency of AI systems, but also broadens their versatility. 
As a transformative concept, neuro-symbolic AI harbors considerable potential to accelerate the quest towards authentic artificial general intelligence.

\begin{itemize}
    \item In the operational dynamics of neuro-symbolic AI, deep learning models primarily employ vectors and matrices. These models utilize multi-layered neural networks to handle input data, each layer manipulating the input through a sequence of vector-matrix multiplications (VMMs). Such transformations allow the network to discern intricate patterns from data, enabling it to carry out tasks like image-based object recognition or text-based sentiment analysis.

    \item The symbolic AI facet of neuro-symbolic AI predominantly operates on rule-based models that adhere to explicit rules, implement search algorithms, and conduct logical inferences. This imparts them a competitive edge in tasks demanding explicit reasoning, knowledge representation, and interpretability. 
    The core component of a symbolic AI model involves decision matching and manipulation, which can be processed through search-based operations. Hyperdimensional computing is one of the symbolic models with significant efficiency and robustness \cite{kanerva2009hyperdimensional,zou2021scalable} and is adopted in this work.  \end{itemize}


Neurosymbolic AI combines two essential components: the "neuro" part, which handles things such as feature extraction or latent space generation through layers of neurons performing vector-matrix multiplication (often accelerated by crossbar arrays), and the "symbolic" part, based on associative search (sped up by content addressable memory (CAM)). The resource allocated between these two components varies, depending on the application, with some tasks demanding a larger neuro part, while others rely more on the symbolic part.
For example, in reasoning tasks, the symbolic and neuro parts often require comparable resources, whereas image classification leans more on the neuro side due to the significance of reliable feature extraction and image segmentation. This inherent trade-off poses a challenge in developing a single hardware accelerator that can efficiently accommodate a wide range of tasks with diverse resource requirements for each component. To address this issue, we introduce an innovative hardware platform that can dynamically switch between functioning as a crossbar array and as CAM blocks. This adaptability allows a single piece of hardware to accommodate different neurosymbolic AI models by dynamically adjusting the architecture to fine-tune resource allocation, ensuring optimal utilization and enhanced performance without unnecessary waste.

To build a hardware accelerator for the neuro-symbolic AI, compute-in-memory (CiM) is a highly promising approach where the network weights are stored in the memory arrays and the computations are performed in analog domain by exploiting the in-situ array operations. 
In this way, frequent and expensive data movement between the computing and memory is avoided,  rendering high-performance and energy-efficient computing. 
Generally, CiM can be done either in current domain or charge domain, where in the former, current of cells in a column are directly summed together. In the latter case, the local computing result within a cell is represented as capacitor charge, which together with other capacitors within a column share the charge to perform the analog summation. 
As shown in Fig.\ref{fig:fig1_overview}(b), when summing up the cell currents, the computing results in current domain are directly impacted by the cell conductance variation, making it challenging in achieving accurate CiM. 
On the other hand, for charge-domain CiM, the capacitor will be charged to fixed voltages corresponding to the computing results of the local input and weight. 
Since these fixed voltages are supplied externally, the accumulated capacitor charge will be weakly dependent on the actual cell current, 
and hence also on its variation. 
With well controlled capacitor variation, the overall computation results can be more accurate compared to the current domain CiM, as illustrated in Fig.\ref{fig:fig1_overview}(c), and thus is the focus of this work. 

\newpage
\begin{figurehere}
	\centering
	\includegraphics [width=0.9\linewidth]{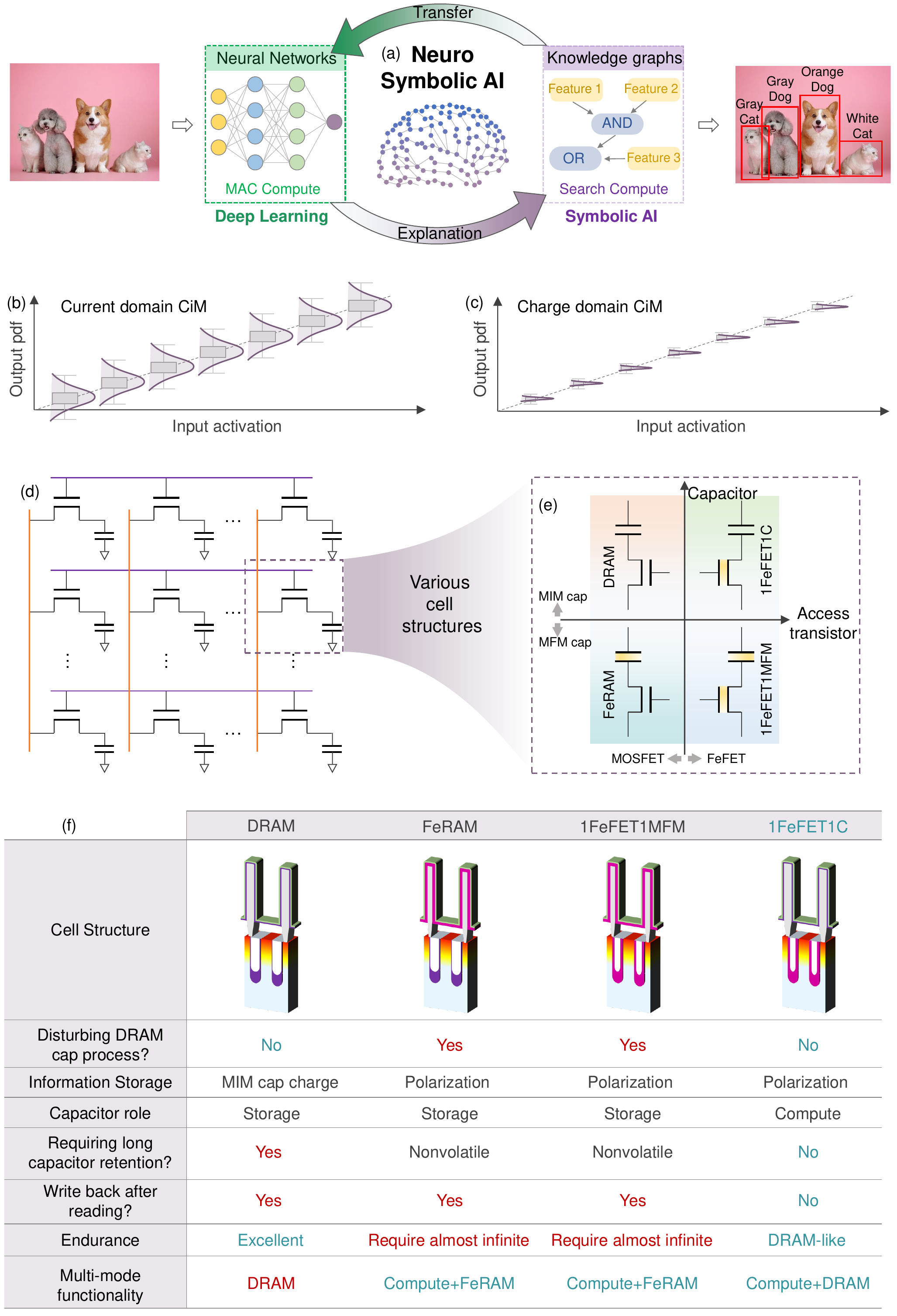}
	\caption{\textit{\textbf{Overview of charge-domain CiM design with different 1 transistor 1 capacitor cells.} (a) The "neuro" and "symbolic" part of neurosymbolic AI. (b) Variation in transistor \textit{V}\textsubscript{TH} and hence ON current results in output error for CiM performed by summing up current in analog domain. (c) Charge domain CiM, where transistor is only used as a switch, shows much better robustness against transistor variation, thus promising for accurate CiM. (d) A typical 1 transistor 1 capacitor array for charge domain CiM. (e) Depending on the implementations of the transistors (i.e., normal transistor or FeFET) and the capacitors (i.e., normal capacitor or MFM capacitor), different unit cells are possible. (f) The proposed work of 1FeFET-1C cell structure is ideal for charge domain CiM due to various appealing inherent features.}}
	\label{fig:fig1_overview}
\end{figurehere}

Charge domain CiM can be implemented with an array of cells, in its simplest structure composed of a transistor and a capacitor cell, as shown in Fig.\ref{fig:fig1_overview}(d).
For this cell, different types of memory, either volatile or nonvolatile, can be exploited, including transistor-type memory or capacitor-type memory. Fig.\ref{fig:fig1_overview}(e) shows the taxonomy of the cell structures for the charge-domain CiM. For example, the classical dynamic random access memory (DRAM) cell is composed of a volatile transistor and a capacitor, where the information is stored temporarily on the capacitor. If the volatile capacitor in a DRAM cell is replaced with a nonvolatile one, such as the ferroelectric capacitor, a nonvolatile ferroelectric random access memory (FeRAM) cell can be realized. 
The discovery of ferroelectricity in CMOS-compatible and highly scalable doped HfO\textsubscript{2} has brought unprecedented opportunities in realizing energy-efficient, logic-compatible, and 3D stackable FeRAM \cite{ramaswamy2023nvdram}. Another set of cells, in parallel to DRAM and FeRAM, replace the volatile transistor within their cell structure with a nonvolatile transistor, as shown in Fig.\ref{fig:fig1_overview}(e). The nonvolatile transistor can be any three-terminal transistor with built-in memory, such as flash transistor or ferroelectric field effect transistor (FeFET). 
Among various implementations, HfO\textsubscript{2}-based FeFET stands out due to its scalability, high performance regarding its low operation voltage and high speed (i.e., 3-4 V, 10ns compared with more than 10 V, $>$100$\mu$s for flash), and cost effectiveness due to ease of integration (i.e., minimum interruption to the CMOS process unlike flash). 
With a FeFET as the transistor and a normal capacitor, as shown in Fig.\ref{fig:fig1_overview}(e), a 1FeFET-1C cell is realized, which will later be shown as a highly promising structure to alleviate the bottlenecks of DRAM in enabling CiM. 
At last, it is also possible to replace the FeRAM cell transistor with a FeFET to realize a 1FeFET-1MFM cell, whose multi-mode functionalities are of high interest and will be explored in the future.

Among the four charge-domain CiM cell structures, the 1FeFET-1C cell is the most promising, as summarized in Fig.\ref{fig:fig1_overview}(f). First, unlike the FeRAM and 1FeFET-1MFM, where the complex DRAM capacitor integration has to be modified by incorporating ferroelectrics, the 1FeFET-1C design keeps the capacitor intact and only changes the DRAM access transistor to FeFET, thus making it easy to exploit the well-established DRAM technology. 
Second, with the 1FeFET-1C cell, information is stored in the nonvolatile FeFET, instead of as free charge in DRAM or polarization in FeRAM. The capacitor in the 1FeFET-1C cell is only dedicated to computation, not weight storage in the CiM array, thus eliminating the periodic refresh needed for DRAM. It is sufficient if the capacitor can hold an intermediate compute result for 10s of ns. This design paves the way for charge-domain CiM accelerators, which have long been hindered by short DRAM retention times. 
Third, a key advantage of 1FeFET-1C CiM is its nondestructive operation. Due to the destructive sensing of DRAM and FeRAM, a parallel activation of an array for analog summation will result in destroy of all the stored information, thus requiring write back of a whole array which kills all the benefits acquired by CiM. In contrast, for the 1FeFET-1C cell, information is stored in the FeFET, which, when multiplied with the input, is reflected as charge temporarily stored in the capacitor, as will be explained in Fig.\ref{fig:fig2_principles}. Though later analog summation will destroy the capacitor charge, 
the capacitors are meant to be reset after the computation is completed anyway. 
Therefore, parallel charge-domain analog computing can be fully exploited using the 1FeFET-1C CiM array, without the need of write back.
Given the properties brought by the unique cell structure, the proposed 1FeFET-1C solution is a remedy to CiM implementation in DRAM, opening door for robust, energy-efficient, and high density CiM hardware.

\section*{\textcolor{sared}{Operation Principles of 1FeFET-1C CiM}}

\label{sec:principle}

The discovery of ferroelectricity in doped HfO\textsubscript{2} \cite{boscke2011ferroelectricity} has triggered significant interests in its integration in FeFET for high-density and energy-efficient embedded NVM. 
By applying a positive/negative pulse on the gate, the ferroelectric polarization is programmed to point toward the channel/gate electrode, setting the FeFET threshold voltage, \textit{V}\textsubscript{TH}, to be low/high state, respectively. 
The stored memory state can then be read through the channel current by applying a read gate bias between the low-\textit{V}\textsubscript{TH} (LVT)  and high-\textit{V}\textsubscript{TH} (HVT) states. 
With its electric field driven write mechanism, FeFET exhibits superior write energy efficiency, thus appealing for CiM applications. With FeFET replacing the access transistor in the DRAM cell, the information can then be stored in a FeFET, alleviating the whole cell design for  ultra-low leakage. 
With such a cell technology, Fig.\ref{fig:fig2_principles}(a) shows the overall structure for the CiM chip for neuro-symbolic AI, where each processing element contains the 1FeFET-1C array core (as shown in Fig.\ref{fig:fig2_principles}(b)) and peripheral circuitry. 
Exploiting the decades-long DRAM development, dense 6F\textsuperscript{2} and 4F\textsuperscript{2} 1FeFET-1C array can be designed, as illustrated in Fig.\ref{fig:fig2_principles}(c). Fig.\ref{fig:sup_tcad} shows the technology computer aided design (TCAD) simulations of recessed FinFET access transistor  with ferroelectric replacing the gate dielectric layer. 
It shows a decent window in this device structure, which can support the CiM operations. 
The 4F\textsuperscript{2} design with vertical gate-all-around access transistor is on the horizon and is the densest design for 2D array. Besides, with 2D DRAM scaling approaching an end, the whole community is witnessing a transition to monolithic 3D DRAM. 
The 1FeFET-1C design is also compatible with the 3D DRAM platform, as seen in the Fig.\ref{fig:sup_3d_structure}, where both the sequential 3D and parallel 3D structures can be considered. Therefore, there is a huge potential for the 1FeFET-1C design to enable dense and efficient CiM hardware.   

\begin{figurehere}
	\centering
	\includegraphics [width=1\linewidth]{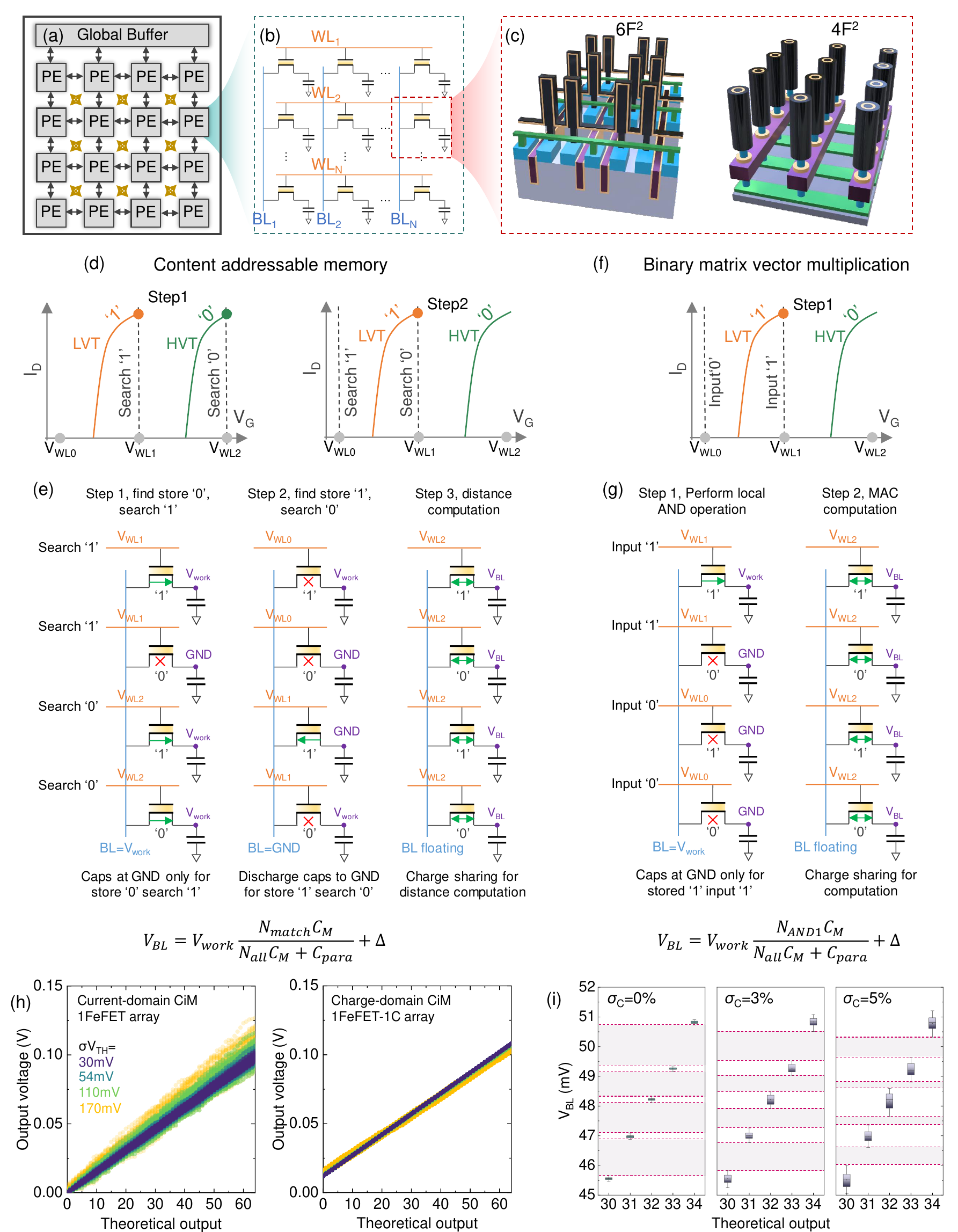}
	\caption{\textit{ \textbf{Operation principles of 1FeFET-1C charge domain computing.}
        (a) The CiM chip consists of multiple PEs and global buffer. 
        (b) The diagram of 1FeFET-1C array core in PE. 
        (c) Dense 6F\textsuperscript{2} and 4F\textsuperscript{2} design of 1FeFET-1C cell, similar to DRAM. 
        The I\textsubscript{D}-V\textsubscript{G} curve of FeFET with LVT/HVT states and their relative relationship with V\textsubscript{WL}s representing different inputs of (d) 2 steps in CAM mode  and (f) 1 step in MAC operation. 
        The detailed computation operations consist of (e) 3 steps in CAM mode and (g) 2 steps in MAC operation for all 4 cases. Their output is linear to the number of charged cells respectively, as shown in the formulas. 
        (h) The relationship between theoretical output code and practical output voltage of current domain 1FeFET array and charge domain 1FeFET-1C array under varying FeFET V\textsubscript{TH} variations. 
        (I) The distribution of practical V\textsubscript{BL} in the worst theoretical output codes under fixed FeFET V\textsubscript{TH} variation and varying capacitor variations.
        }}
	\label{fig:fig2_principles}
\end{figurehere}

The principles of 1FeFET-1C based CiM operation are illustrated in Fig.\ref{fig:fig2_principles}. 
The FeFET \textit{V}\textsubscript{TH} represents the weight, where LVT/HVT state indicates '1'/'0', 
respectively.
The voltage applied at the word line (WL) connected to the FeFET gate represents the input. 
The proposed structure supports  both associative search and VMM
functions, corresponding to logical XNOR and AND operations at each local cell.
Starting from fully discharged capacitors, the parallel associative search operation in the CAM mode is realized via a two-step capacitor charging/discharging process and a final charge-sharing step for computation,
while the multiply-accumulation (MAC) operation of VMM involves one-step charging and one step  charge sharing.  
In the CAM mode, three input  voltage levels at WLs, i.e., \textit{V}\textsubscript{WL1}, \textit{V}\textsubscript{WL2} and \textit{V}\textsubscript{WL3},  sandwich two \textit{V}\textsubscript{TH}s,
as illustrated in Fig.\ref{fig:fig2_principles}(d). Fig.\ref{fig:fig2_principles}(e) shows the three steps needed to complete one associative search operation. 
First, BL of the cells  is raised to \textit{V}\textsubscript{work}, and WL is set to \textit{V}\textsubscript{WL1}/\textit{V}\textsubscript{WL2} representing searching '1'/'0', respectively.
In this way, only the capacitors in the cells storing '0' are not charged when searching '1', as that is the only case where FeFET with the  HVT state is cut-off with \textit{V}\textsubscript{WL1} applied to the gate.
In the second step, BL is connected to GND and WL is set to \textit{V}\textsubscript{WL0}/\textit{V}\textsubscript{WL1} representing searching '1'/'0', respectively. 
Similarly, in this case, only capacitors in the cells storing '1' are discharged to ground when searching '0', as only the FeFET with the LVT state (i.e., storing '1') in those cells is turned ON 
with \textit{V}\textsubscript{WL1} higher than the LVT state applied to the gate.
By performing these two steps, the capacitors of the matched cells are charged to $V_{work}$, while those of the mismatched cells are discharged to ground, indicating the completion of  XNOR operation at the cell level. 
Finally, BL is left floating, and WL is raised to \textit{V}\textsubscript{WL2} to activate all cells within a column for charge sharing on the BL. 
The final voltage of BL is linearly proportional to the number of matched cells within a column, representing the Hamming distance (HD) between the search query and stored vectors. It is expressed as
\begin{equation}
\begin{split}
    V_{BL} &= V_{work}\frac{N_{match}C_M}{N_{all}*C_M+C_{para}}+\Delta \\
    &=V_{work}\frac{C_M (N_{all}-\sum\limits_{n=1}^{N_{all}} Search_n \oplus Stored_n)  }{N_{all}*C_M+C_{para}}+\Delta \\
    &=V_{work}\frac{C_M (N_{all}-HD(Search, Stored))  }{N_{all}*C_M+C_{para}}+\Delta
\end{split}
\end{equation}
where  $N_{match}$ and $N_{all}$ are the number of matched cells 
and the total cells in a column, respectively. $C_M$ and $C_{para}$ are the equivalent cell capacitance and BL parasitic capacitance, respectively. 
$\Delta$ represents the negligible remnant charges on the mismatched cells.
Part of $\Delta$ is due to charge injection when WL is raised, which is why \textit{V}\textsubscript{BL} at theoretical output 0 is higher than 0V. Note that after charge sharing, the BL can be grounded to discharge all the capacitors for initialization of the next computing step.

Similar charge and discharge steps can be performed in charge domain for MAC operations, given the input voltage and stored state configuration shown in  Fig.\ref{fig:fig2_principles}(f). 
Fig.\ref{fig:fig2_principles}(g) depicts our proposed 2-step MAC operations. 
In step 1, BL of the array column is raised to \textit{V}\textsubscript{work}, and WL is set to \textit{V}\textsubscript{WL1}/\textit{V}\textsubscript{WL0} representing input '1'/'0', respectively.
Only cells that store '1' (i.e., at the LVT state) and receive an input '1' (i.e., \textit{V}\textsubscript{WL1} applied at gate) are charged to \textit{V}\textsubscript{work},
representing the bit-wise AND operation. 
In step 2, the capacitor charges  within the column are shared by floating the BL and setting all WLs to \textit{V}\textsubscript{WL2}.
This step leads to the accumulation of bit-wise AND results in the charge domain. 
The final   BL voltage is linearly proportional to the number of cells with the multiplication result of '1', representing the MAC result between the input vector and stored weights. 
It is expressed as
\begin{equation}
\begin{split}
    V_{BL} &= V_{work}\frac{N_{AND1}C_M}{N_{all}*C_M+C_{para}}+\Delta \\
    &=V_{work}\frac{C_M \sum\limits_{n=1}^{N_{all}} Input_n \times Stored_n  }{N_{all}*C_M+C_{para}}+\Delta
\end{split}
\end{equation}
where the $N_{AND1}$ is the number of cells in a column that gives an output of logical '1'. 
From the operations, it is clear that FeFET only acts as a nonvolatile switch and its large ON/OFF ratio is exploited to ensure full swing charge and discharge.
Therefore,
compared to current domain CiM designs, our proposed 1FeFET-1C charge domain CiM  exhibits superior immunity to FeFET \textit{V}\textsubscript{TH} variation. 

Using a calibrated FeFET compact model \cite{ni2018circuit} and 45nm predictive technology model (PTM) \cite{zhao2006new},   the 1FeFET-1C CiM array operations are validated in SPICE via Monte Carlo simulations. 
Fig.\ref{fig:fig2_principles}(h) shows the relationship between the array output voltage distribution and theoretical output for both 1FeFET current domain CiM  and 1FeFET-1C charge domain CiM, considering varying degrees of FeFET \textit{V}\textsubscript{TH} variation. 
The 1FeFET-1C charge domain CiM provides much more accurate output results, therefore, demonstrating great robustness and  linearity. 
In addition to FeFET \textit{V}\textsubscript{TH} variation, the capacitor variation also affects the computation accuracy, 
and the relationship between \textit{V}\textsubscript{BL}, the varying capacitance, and the proportion of activated cells during the operation can be expressed as: 
\begin{equation}
\label{equ: VBL_var}
    V_{BL} = V_{work}\frac{\sum\limits_{n=1}^{N_{all}}v_i C_{M_i}}{\sum\limits_{n=1}^{N_{all}}C_{M_i}+C_{para}}
\vspace{0ex}
\end{equation}
where $v_i$ follows a Bernoulli distribution based on the proportion of activated cells, and $C_{M_i}$ follows a Gaussian distribution.
Extensive sampling reveals that the standard deviation $\sigma_{V_{BL}}$ is maximized when the proportion of activated cells is 0.5. 
Considering both FeFET \textit{V}\textsubscript{TH} variation and varying capacitor variation, the distribution of \textit{V}\textsubscript{BL} under the worst case is illustrated in Fig.\ref{fig:fig2_principles}(i). Even in scenarios with $\sigma_{C_M}$ = 5\% and $\sigma_{Vth} = 54 mV$, the 1FeFET-1C design ensures a sufficient sense margin. 

\section*{\textcolor{sared}{Experimental Demonstration of 1FeFET-1C CiM}}
\label{sec:experiment}

Experimental validation is also performed to verify the operational  linearity of charge-domain computing with fabricated 1FeFET-1C array. 
FeFET devices were fabricated using a 28 nm node gate-first high-k metal gate CMOS process on 300 mm silicon wafers. Detailed process information can be found in \cite{trentzsch201628nm}. 
Fig.\ref{fig:fig3_exp}(a) shows the fabricated FeFET featuring a gate stack composed of a poly-crystalline Si/TiN/Si-doped HfO\textsubscript{2} (8 nm)/SiO\textsubscript{2} (1 nm)/p-Si, as shown in Fig.\ref{fig:fig3_exp}(b).
Detailed integration process is illustrated in the "Materials and Methods" section.
Fig.\ref{fig:fig3_exp}(c) shows the FeFET \textit{I}\textsubscript{D}-\textit{V}\textsubscript{G} characteristics for the LVT and HVT states defined with +4V and -4V, 1$\mu$s write pulse, respectively. 
The device exhibits a large memory window ($\sim$1V) and well-tempered device-to-device variation with W/L=500nm/500nm. Faster programming can also be achieved in FeFET, as indicated in Fig.\ref{fig:fig3_exp}(d), where below 100ns programming can be realized. 
These device properties demonstrate that FeFET is a highly promising technology platform.

Then cell-level and array-level demonstrations are performed. For the ease of integration, only planar devices are used in this work, where planar FeFET and planar metal-insulator-metal (MIM) capacitors are adopted. Though it is not the same as the dense DRAM array shown in Fig.\ref{fig:fig2_principles}(c), it is good enough to validate the working principles. Fig.\ref{fig:fig3_exp}(e) and (f) shows the charging transients for the 1FeFET-1C cell at the LVT and HVT states, respectively. 
When \textit{V}\textsubscript{WL}=1V is applied to the gate of FeFET storing the LVT state, it can reliably pass BL voltage up to 0.5V. 
Higher \textit{V}\textsubscript{BL} can incur \textit{V}\textsubscript{TH} loss. 
In contrast, for the cell at the HVT state, the FeFET in the cell is cut-off, causing no charging of the capacitor. These cell results clearly validate that the large ON/OFF ratio of FeFET can enable successful operation. For the array level demonstration, an 8x1 column is adopted, as shown in Fig.\ref{fig:fig3_exp}(g), where the rightmost layout shows the 8 FeFETs and 8 capacitors used. 
The corresponding implemented chip is also presented on the left. 
With the array, the operations illustrated in Fig.\ref{fig:fig2_principles} are validated, where the results are shown in Fig.\ref{fig:fig3_exp}(h). 
It illustrates the \textit{V}\textsubscript{BL} transients during the charge sharing step where the computation happens on the floated BL. 
It clearly shows that the larger the theoretical output, the higher the \textit{V}\textsubscript{BL}. The peak {V}\textsubscript{BL} exhibits a linear dependence on the theoretical output, as summarized in Fig.\ref{fig:fig3_exp}(i). Therefore, these results validate the proposed working principles of the 1FeFET-1C charge domain CiM.    

\begin{figurehere}
	\centering
	\includegraphics [width=1\linewidth]{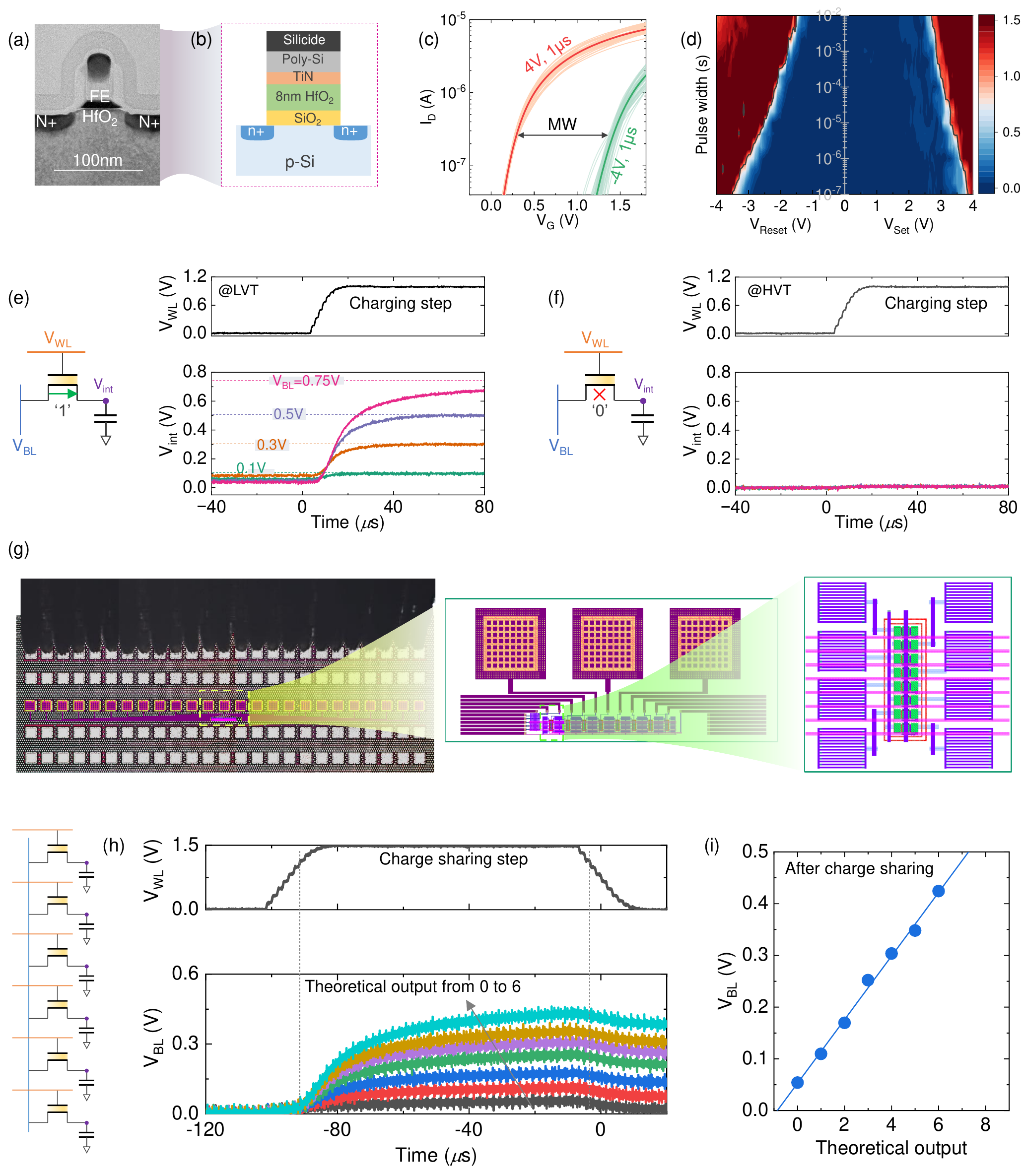}
	\caption{\textit{\textbf{Experimental demonstration of charge-domain 1FeFET-1C CiM design.} (a) Cross section of a fabricated FeFET device. (b) Schematic of FeFET gate stack. (c) FeFET \textit{I}\textsubscript{D}-\textit{V}\textsubscript{G} characteristics for the LVT and HVT states, exhibiting 1V memory window and well-tempered device variation. (d) The switching dynamics of FeFET device shows that fast programming can be achieved. The heat map represents the memory window, and the boundary line indicates a memory window of 1V. Left half dynamics are obtained by initializing the device with 4V and applying the negative erase voltage \textit{V}\textsubscript{Reset}, while the right half is by initializing with -4V and applying positive program voltage \textit{V}\textsubscript{Set}. Experimental charging transients for the 1FeFET-1C cell at (e) the LVT and (f) the HVT states clearly validate the successful cell operation due to the large ON/OFF ratio of FeFET. (g) The fabricated chip layout with a zoom-in view of 8 FeFETs and 8 capacitors. (h) Experimental charge sharing transients for an 8x1 column clearly shows that the Bl voltage corresponding to the final computation result aligns with the theoretical output, and (i) the peak BL voltage is linearly proportional to the theoretical output, sufficiently validating the proposed working principles of the 1FeFET-1C CiM design. }}
	\label{fig:fig3_exp}
\end{figurehere}

\section*{\textcolor{sared}{Benchmarking and Application of 1FeFET-1C CiM Array}}
\label{sec:benchmarking_application}

To demonstrate the advantages of the proposed 1FeFET-1C charge domain CiM design, we evaluated and compared several state-of-the-art charge-domain CiM designs, including those based on SRAM \cite{valavi201964}, RRAM \cite{jeong2022variation}, MRAM \cite{jacob2023nonvolatile}, and FeFET \cite{yin2021enabling}, respectively, all based on  SPICE simulations for fair comparisons.
For the evaluation, calibrated FeFET models are adopted \cite{ni2018circuit}. 
The MAC operations over an array with a word length of 64 are performed and evaluated in terms of energy and computation latency. 
As shown in Fig.\ref{fig:fig4_eval}(a), 1FeFET-1C achieves the lowest energy-delay product (EDP) across all designs. 
Although the two-step operation results in a longer latency, this is compensated by the lower energy consumption, due to the low \textit{V}\textsubscript{work}. 
Fig.\ref{fig:fig4_eval}(b) compares the cell areas in terms of feature size (F). 
Benefiting from the DRAM-like structure, 1FeFET-1C can achieve a minimum cell area of 6F\textsuperscript{2}. 
With the future monolithically integrated 3D 1T-1C DRAM on the horizon as illustrated in Fig.\ref{fig:sup_3d_structure}, a high density 1FeFET-1C can be expected.  
Fig.\ref{fig:fig4_eval}(c) further investigates the latency and energy of the proposed 1FeFET-1C array core in performing computation with varying array sizes. Peripheral ADCs are not included for this evaluation. Fig.\ref{fig:sup_array_wt_wo_ADC}(b) shows the similar evaluation, but with the successive approximation ADC (SAR ADC) at the columns included. Evaluating both gives a holistic picture of the 1FeFET-1C performance.
As the number of rows (\textit{N}) increases, given the constant \textit{V}\textsubscript{work}, the charge-sharing scheme reduces the sense margin of output voltage, demanding higher precision of the sense amplifier. 
To address this issue for large \textit{N}, \textit{V}\textsubscript{work} needs to increase to maintain a sufficient output sense margin.
For \textit{N} less than 64, with a fixed \textit{V}\textsubscript{work}, the energy scales linearly with \textit{N}, and the latency remains constant. 
When \textit{N} exceeds 64 and \textit{V}\textsubscript{work} increases, the energy consumption approximately scales quadratically with \textit{N}. 
A higher \textit{V}\textsubscript{work} increases the charging and discharging time of cells, leading to longer latency as \textit{N} grows. 
Regarding the number of columns (\textit{M}), since each column operates independently in parallel, the energy consumption has a linear relationship with \textit{M}, while the latency remains unaffected. If ADCs with resolution of $log_2(N)$ are included, more columns means the need of multiplexing, which also increases the latency, as shown in Fig.\ref{fig:sup_array_wt_wo_ADC}(b).

With the 1FeFET-1C array's performance metrics evaluated, its impact on the overall neuro-symbolic AI systems is studied. 
The RAVEN dataset \cite{zhang2019raven}, 
a benchmark designed to evaluate AI models’ abilities in relational and analogical visual reasoning, is adopted for the evaluation, as it requires models to solve visual puzzles by recognizing patterns in attributes like shape, size, and position, which involves both relational reasoning and pattern matching. 
Fig.\ref{fig:fig4_eval}(d) shows the overall system framework, composed of a  neural network-based model for feature extraction and a brain-inspired hyperdimensional computing (HDC) model for symbolic component to support reasoning \cite{hersche2023neuro}.
Neural networks excel at recognizing patterns,
but they lack explicit reasoning abilities, which HDC excels through high-dimensional vector
encoding and manipulation.
HDC model replicates various functions of the human brain within its algebraic framework, including memory, associative learning, sequential order and associative search.
It allows for the binding of concepts into composite structures, akin to how the human brain associates ideas and contexts, thereby enabling logical and relational reasoning. 
This is particularly essential in tasks requiring the understanding of relationships, analogies, and sequences, as seen in the RAVEN dataset. 
In this framework, the neural network part is executed once to generate symbolic representations from the input data. 
These representations, which
capture essential features and patterns, are then handed off to the HDC model. Since the symbolic representations are abstract and versatile, they can be reused by the HDC component across multiple reasoning tasks without needing to rerun the neural network. 
While this ability to reuse symbolic representations is not explicitly present in the RAVEN dataset, it is a fundamental aspect of human cognition. For example, once we learn the concept of "symmetry," we can apply it across various contexts—whether identifying symmetry in a piece of art, understanding balance in physical structures, or even in everyday tasks like arranging objects on a shelf. Similarly, the understanding of a pattern like "cause and effect" allows us to reason about situations ranging from mechanical systems to social interactions without needing to relearn the concept each time. For our proposed 1FeFET-1C array that implements both the neural network and the HDC components, the symbolic (HDC) part will be engaged more frequently, enabling continuous and dynamic reasoning while the neural network part remains idle until new data inputs are required. With such an architecture, complex visual reasoning tasks in RAVEN can be handled with much better accuracy and interpretability than purely reasoning neural models like wild rational network (WReN) \cite{hersche2023neuro}.

\begin{figurehere}
	\centering
	\includegraphics [width=1\linewidth]{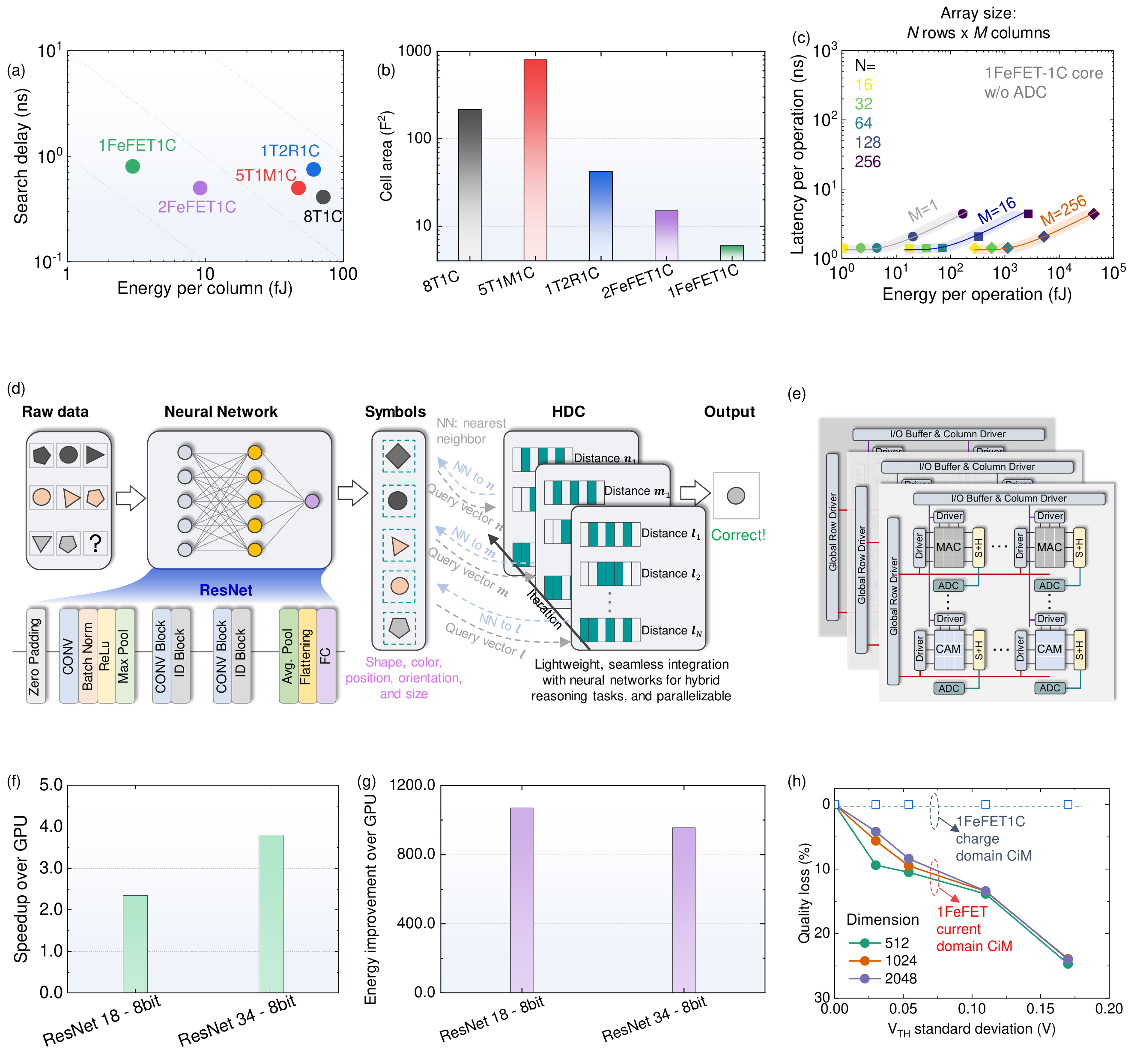}
    \vspace{-6ex}
	\caption{\textit{ \textbf{Evaluation and benchmarking of 1FeFET-1C array for neuro-symbolic AI applications.} (a) The latency vs. energy to execute a MAC operation for a column with 64 cells with different charge-domain cell designs. (b) The corresponding cell area. Our proposed 1FeFET-1C cell design exhibits the best energy-delay product and density. (c) The array operation delay vs. energy for the 1FeFET-1C array core without including ADC at different array sizes. (d) Neuro-symbolic AI model architectures for the evaluation of the Raven dataset that includes a ResNet for the neural network part and HDC for the symbolic part. (e) The overall system architecture that includes the MAC part for symbol extraction and the CAM part for HDC. (f) The speed up and (g) the energy improvement over the GPU implementation of the HDC based symbolic component. (h) The quality loss of the overall system when considering FeFET }V\textit{\textsubscript{TH} variation for both the 1FeFET based current domain CiM and the 1FeFET-1C based charge domain CiM.}}
	\label{fig:fig4_eval}
\end{figurehere}

Given that our proposed 1FeFET-1C design can be configured to function both MAC and associative search, the neuro-symbolic system can be easily mapped to our 1FeFET-1C CiM based hardware, as illustrated in Fig.\ref{fig:fig4_eval}(e). Again, as the same 1FeFET-1C array can realize the MAC and associative search function, it allows dynamic resource allocation between the two components, thus capable of supporting various neuro-symbolic AI tasks with varying demands.
With this hardware configuration, we simulate and compare the system's latency and energy consumption when performing the RAVEN tasks with the baseline NVIDIA A6000 GPU. The code has been optimized to ensure over 95\% GPU utilization on average. 
In our neuro-symbolic architecture, the front-end neural network model operates with an 8-bit representation, while the HDC backbone functions using a 1-bit representation. This reduction enables memory and energy savings while preserving nearly identical accuracy levels to the original architecture. 
The neural network is supported by multiple cycle operations using the crossbar block for MAC with 1-bit FeFET devices, and the HDC operation is implemented using the CAM block we introduced.
Fig.\ref{fig:fig4_eval}(f) and (g) shows the speedup and energy improvement over the GPU when the ResNet-18 and ResNet-34 are used as neural network for the feature extraction, respectively. 
It shows 2.5$\times$ and 4$\times$ speedup for ResNet-18 and ResNet-34, while the energy improvement of 5,000$\times$ and around 4,000$\times$, respectively, over the baseline GPU. Therefore, the proposed design exhibits superior energy-delay product improvement over the GPU, demonstrating great promise for efficient reasoning tasks.

Since the neuro-symbolic framework's reasoning tasks demand precise similarity measures, the CiM array based hardware may degrade performance given the variations of FeFETs.
Therefore the impact of these cell variations at system level is evaluated, assuming varying FeFET \textit{V}\textsubscript{TH} variations $\sigma_{Vth}$ at
30mV, 54mV, 110mV, and 170mV. 
For the evaluation, the current-domain  1FeFET CiM array and charge-domain  1FeFET-1C CiM array are compared. 
The neuro-symbolic reasoning tasks are compiled and executed across various symbolic vector dimensionalities including 512, 1024, and 2048. 
The final similarity values are computed by mapping the range of values affected by noise into a probability mass function (PMF), 
which is used to model the likelihood of different high-dimensional vector representations, aiding in probabilistic reasoning. By computing the PMF, the model can assign probabilities to symbolic representations, which helps in selecting the most likely reasoning paths or outcomes based on the encoded information.
The RAVEN dataset's rigorous demands provides an ideal testbed to assess both the circuit's operational fidelity and the task performance degradation caused by hardware variations.
The quality loss, defined as the difference in accuracy between ideal and practical conditions with FeFET variations, was measured. 
For the current-domain CiM, with $\sigma_{Vth}$=30mV, the quality loss was relatively minimal, with a degradation of 9.4\% at a dimension of 512, which decreases to 4.2\% at a higher dimension of 2048. 
As expected, the higher the FeFET variation, the more quality loss will be. The most substantial impact is observed at $\sigma_{Vth}$=170mV, where the quality loss soars to 24.7\% at a dimension of 512, though decreases to 17.0\% at 2048. These results indicate a clear trend: higher dimensions generally exhibit a higher resilience to device variation, but as the device variation increases, the accuracy degradation becomes more significant across all dimensionalities.
In contrast, for the proposed 1FeFET-1C charge-domain CiM, there is no noticeable quality loss at a 170mV $\sigma_{Vth}$ as no precise ON state conductance is required in this scheme as long as a sufficient ON/OFF ratio of FeFET as the access transistor is exhibited, as clearly demonstrated in Fig.\ref{fig:fig2_principles}(h). Therefore, the proposed 1FeFET-1C solution also exhibits superior robustness to the device variability.  


\section*{\textcolor{sared}{Conclusion}}
\label{sec:conclusion}

We build a hardware accelerator for the neuro-symbolic AI models with  an ultra-compact and scalable ferroelectric CiM design, 
by leveraging its inherent massive parallel, in-memory computing capability and dual-mode functions  that support both the VMM operations of "neuro" part and the associative search operations of "symbolic" part.
We fabricate  the proposed charge-domain CiM design that consists of a FeFET device and a capacitor using a 28 nm node gate-first high-k metal gate CMOS process on 300 mm silicon wafers. 
The 1FeFET-1C structure   enables the most compact, energy efficient CiM demonstrated to date that is suitable for neuro-symbolic AI implementation, especially considering the compatibility to DRAM technology and immunity against multi-bit FeFET device variations.
We theoretically and experimentally demonstrate the ability of the fabricated 1FeFET-1C array to calculate Hamming distances between search vectors and stored entries in a 3-step process, and to calculate the MAC results for neural networks in a 2-step process, both outperforming the state-of-the-art counterpart designs in terms of energy and area efficiency.
We further show that, by configuring the proposed CiM design into either MAC array or CAM, the proposed 1FeFET-1C CiM design is sufficient for acceleration of feature extraction and explicit searching required in a neuro-symbolic AI framework for image classification tasks.   
As such, the 1FeFET-1C CiM design represents a remedy to CiM with DRAM technology, and provides a promising hardware platform towards implementing efficient neuro-symbolic AI that facilitates  the developments towards authentic artificial general intelligence.

\section*{\textcolor{sared}{Materials and Methods}}
\label{sec:materials}


\subsection*{Device fabrication}
FeFETs tested in this work are integrated on the 28 nm industrial high-$\kappa$ metal gate (HKMG) platform. The device has a gate stack composed of a poly-crystalline Si/TiN (2 nm)/doped HfO\textsubscript{2} (8 nm)/SiO\textsubscript{2} (1 nm)/p-Si. The ferroelectric gate stack process starts with an 8 nm thick doped HfO\textsubscript{2} deposition through atomic layer deposition process on the 300 mm silicon wafer, which is covered with a thin SiO\textsubscript{2} interfacial layer. Then a TiN metal gate electrode was deposited using physical vapor deposition (PVD), followed by the poly-Si gate electrode deposition. The source and drain n+ regions were obtained by phosphorous ion implantation, which were then activated by a rapid thermal annealing (RTA) at approximately 1000 $^\circ$C. In addition to the dopant activation, the RTA process will also stabilize the ferroelectric orthorhombic phase within the doped HfO\textsubscript{2}.

\subsection*{Electrical characterization}
The FeFET device characterization was performed with a PXI-Express system from National Instruments, using a PXIe-1095 cassis, NI PXIe-8880 controller, NI PXIe-6570 pin parametric measurement unit (PPMU) and NI PXIe-4143 source measure unit (SMU).
Prior to characterization all FeFETs are preconditioned using the SMUs by cycling them 100 times with the pulses of +4.5 V, -5 V with a pulse length of 500 ns each. Bulk and drain terminals are tied to ground at all times. The read operation takes approximately 7 ms. The FeFETs are then programmed or erased by the application of-4 V or +4 V for 1 µs for erase or program, respectively. The matchline waveform is recorded for all FeFETs being activated by a gate voltage pulse of 1.2V or 1.5V. The matchline voltage is then recorded using a Tektronix oscilloscope. Sequentially the individual FeFETs get programmed. 

\subsection*{Benchmarking of Neuro-Symbolic AI for Reasoning Tasks}
To evaluate the network for the RAVEN dataset in the context of neuro-symbolic AI, we adopted a benchmarking approach designed to measure the model’s effectiveness in handling complex reasoning tasks. The RAVEN dataset, inspired by Raven’s Progressive Matrices, is a benchmark for relational and analogical visual reasoning, where models must recognize patterns in attributes such as shape, size, and position to complete visual puzzles.

Our evaluation involved using a hybrid architecture that combines a ResNet-based neural network for feature extraction and a hyperdimensional computing (HDC) model for symbolic reasoning. The ResNet processes visual inputs and extracts features like shape and size, while the HDC model applies rule-based reasoning to solve the puzzles by comparing and analyzing the extracted features. The system was configured to operate efficiently on our 1FeFET-1C compute-in-memory hardware, leveraging both MAC operations for neural network processing and CAM operations for symbolic reasoning tasks.

The system's performance was benchmarked against the NVIDIA A6000 GPU, optimizing for energy consumption and latency. Additionally, we tested the resilience of the system under different levels of FeFET device variations, evaluating how hardware variability impacts the accuracy and efficiency of solving the RAVEN tasks. This setup provides a comprehensive analysis of the network’s capability in handling complex reasoning tasks while ensuring hardware efficiency and robustness.

\bibliography{ref}

\bibliographystyle{Nature}

\section*{\textcolor{sared}{Acknowledgments}}

\subsection*{Author contributions}
X.Y., T.K., and K.N. proposed and supervised the project. F.M., M.L., R.O., N.L., S.D., and Z.Z. performed the experimental verification of the proposed design. Y,J., J.D., Z.S. and C.Z. conducted SPICE simulations and verification. H.E.B. and M.I. performed system level benchmarking. All authors contributed to write up of the
manuscript.

\subsection*{Competing interests}
\label{sec:conflict}
The authors declare no conflicts of interest.

\subsection*{Data and materials availability}
\label{sec:data}

The data that support the plots within this paper and other findings of this study are
available from the corresponding author on reasonable request.



\newpage
\renewcommand{\thefigure}{S\arabic{figure}}
\renewcommand{\thetable}{S\arabic{table}}

\onecolumn
\centering
\textbf{\Large Supplementary Information}
\setcounter{figure}{0}
\setcounter{table}{0}
\setcounter{page}{1}

\begin{flushleft} 
\textbf{\large Target FeFET Characteristics for Current-Based In-Memory Computing}
\end{flushleft}

\justify
For current based in-memory computing using FeFETs, such as crossbar array for matrix-vector multiplication (multiply and accumulation, MAC) operation (Fig.\ref{fig:figs1_device_for_imc}(a)) and the content addressable memory for associative search (Fig.\ref{fig:figs1_device_for_imc}(b)), it is highly desirable to have a tight distribution of the ON current for FeFETs. For conventional FeFETs, the device current variation is closely related to the transistor \textit{V}\textsubscript{TH} variation. Especially for the ON current, the \textit{V}\textsubscript{TH} variation has an one-to-one correspondence to the \textit{I}\textsubscript{D} variation, as shown in Fig.\ref{fig:figs1_device_for_imc}(c). The ON current variability may induce output overlaps, thus sensing error, which could degrade the operation accuracy.

\begin{figurehere}
	\centering
	\includegraphics [width=0.8\linewidth]{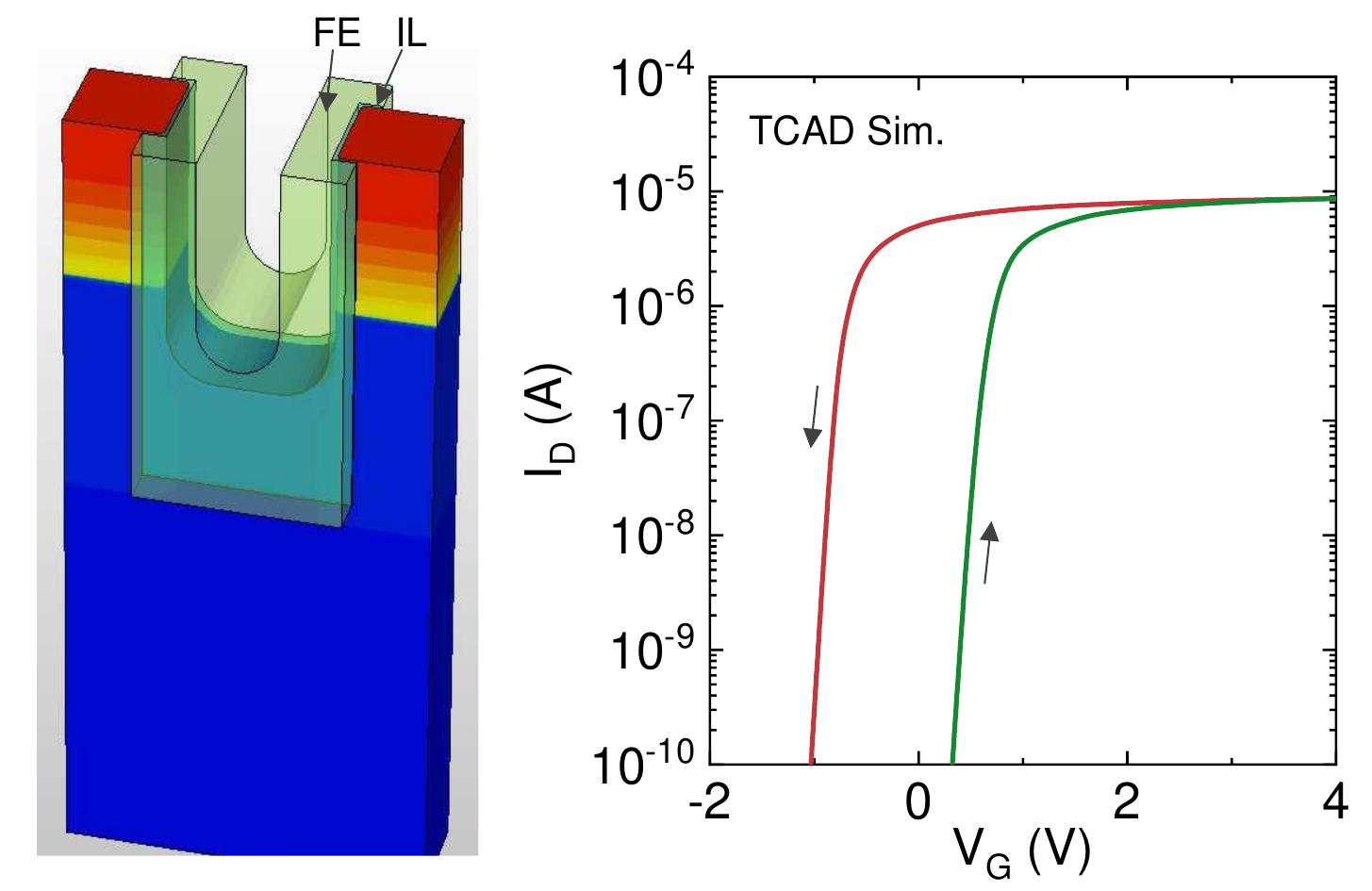}
	\caption{\textit{TCAD simulation of saddle FinFET based FeFET.}}
	\label{fig:sup_tcad}
\end{figurehere}

\newpage
\begin{flushleft} 
\textbf{\large 3D 1FeFET-1C Array}
\end{flushleft}

\justify
The 1FeFET-1C array is also 3D compatible, akin to 3D DRAM. Two typical 3D array structures can be considered, one is the sequential 3D and the other one is a parallel 3D. In the former structure, the 1FeFET-1C tier is fabricated one after another and in the latter, the whole stack is fabricated top-down.

\begin{figurehere}
	\centering
	\includegraphics [width=0.8\linewidth]{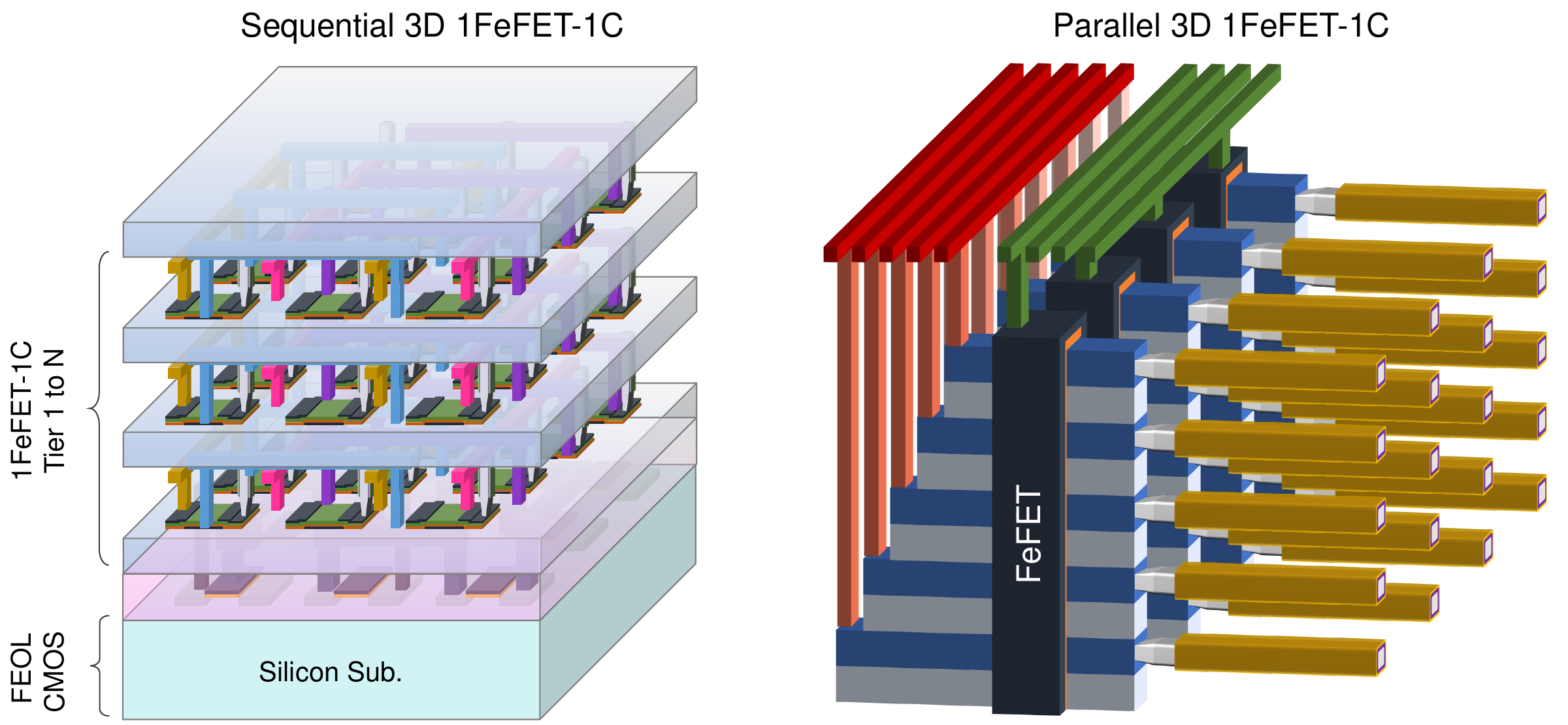}
	\caption{\textit{Sequential and parallel 3D 1FeFET-1C array.}}
	\label{fig:sup_3d_structure}
\end{figurehere}




\newpage
\begin{flushleft} 
\textbf{\large 1FeFET-1C Array Latency and Energy Evaluation}
\end{flushleft}

To evaluate the latency and energy consumption of the proposed 1FeFET-1C array, two scenarios are considered. In the case, just the array core is included, without considering array peripherals, as shown in Fig.\ref{fig:fig4_eval}(c) also replotted here in Fig.\ref{fig:sup_array_wt_wo_ADC}(a). 
The other case is for the array including the successive-approximation ADC (SAR ADC) at the peripherals. SAR ADCs are also designed and evaluated in SPICE. 
When ADCs are included, multiplexing among multiple columns are considered to align  the hardware overhead for ADC with the array, and the performance metrics are shown in Fig.\ref{fig:sup_array_wt_wo_ADC}(b).

\begin{figurehere}
	\centering
	\includegraphics [width=1\linewidth]{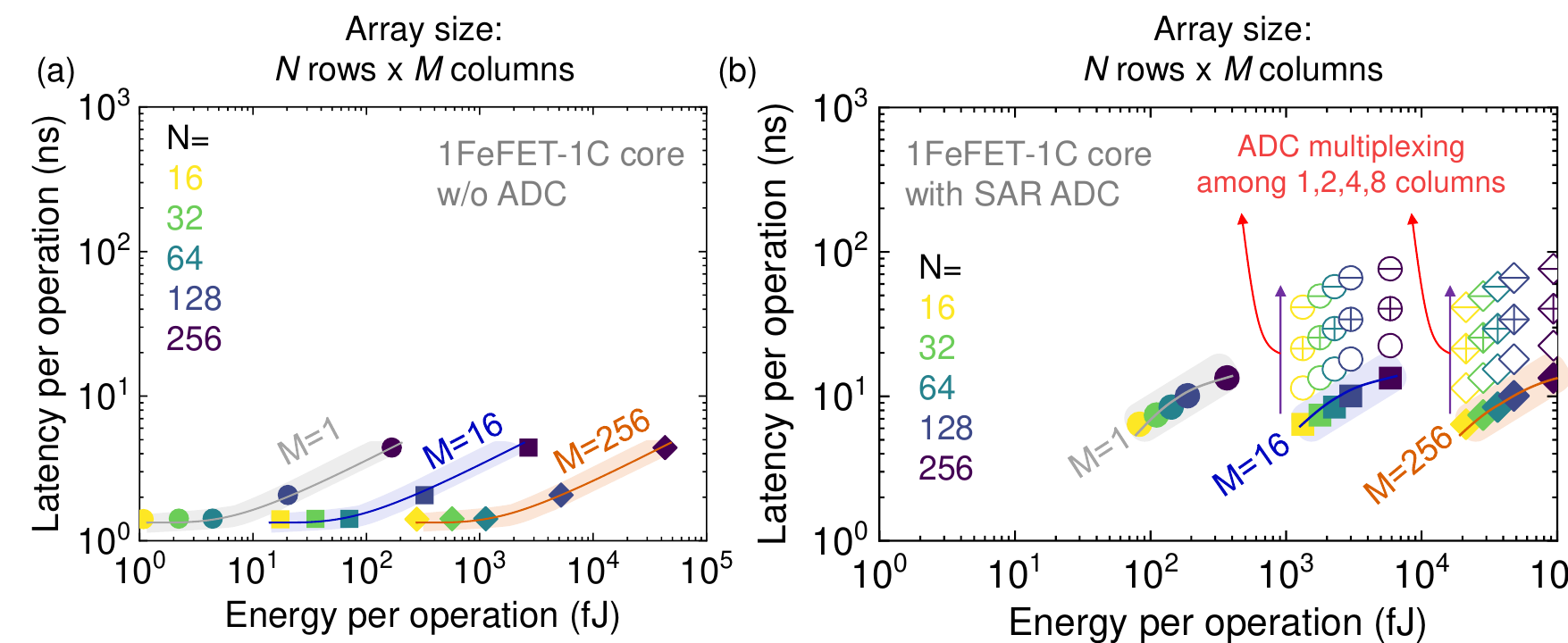}
	\caption{\textit{The array operation delay vs. energy for (a) the 1FeFET-1C array core only and (b) 1FeFET-1C array with SAR ADC at different array sizes.}}
	\label{fig:sup_array_wt_wo_ADC}
\end{figurehere}

\newpage
\begin{flushleft} 
\textbf{\large Experiment Design}
\end{flushleft}

In our experiments, we set out to evaluate the performance and versatility of our design across three distinct image classification tasks, progressively varying in complexity. These tasks were conducted on the Fashion MNIST (FMNIST), CIFAR-10, and CIFAR-100 datasets. Our first step involved fine-tuning the deep neural network (DNN) to extract 512-dimensional features. Notably, the FMNIST dataset required only 20 epochs for training, while the CIFAR-10 and CIFAR-100 datasets demanded a more extensive training period of 200 epochs. To assess the influence of the depth of our feature extraction module, we began with an 18-layer Resnet DNN and systematically reduced it to 25\%, 50\%, and 75\% of the convolutional layers. Concurrently, our hyperdimensional computing (HDC) model underwent training over 20 epochs, with variations in dimensionalities tested, ranging from 512 to 10,000.

\begin{figure*}[t!]
    \centering
    \includegraphics[width=0.85\textwidth]{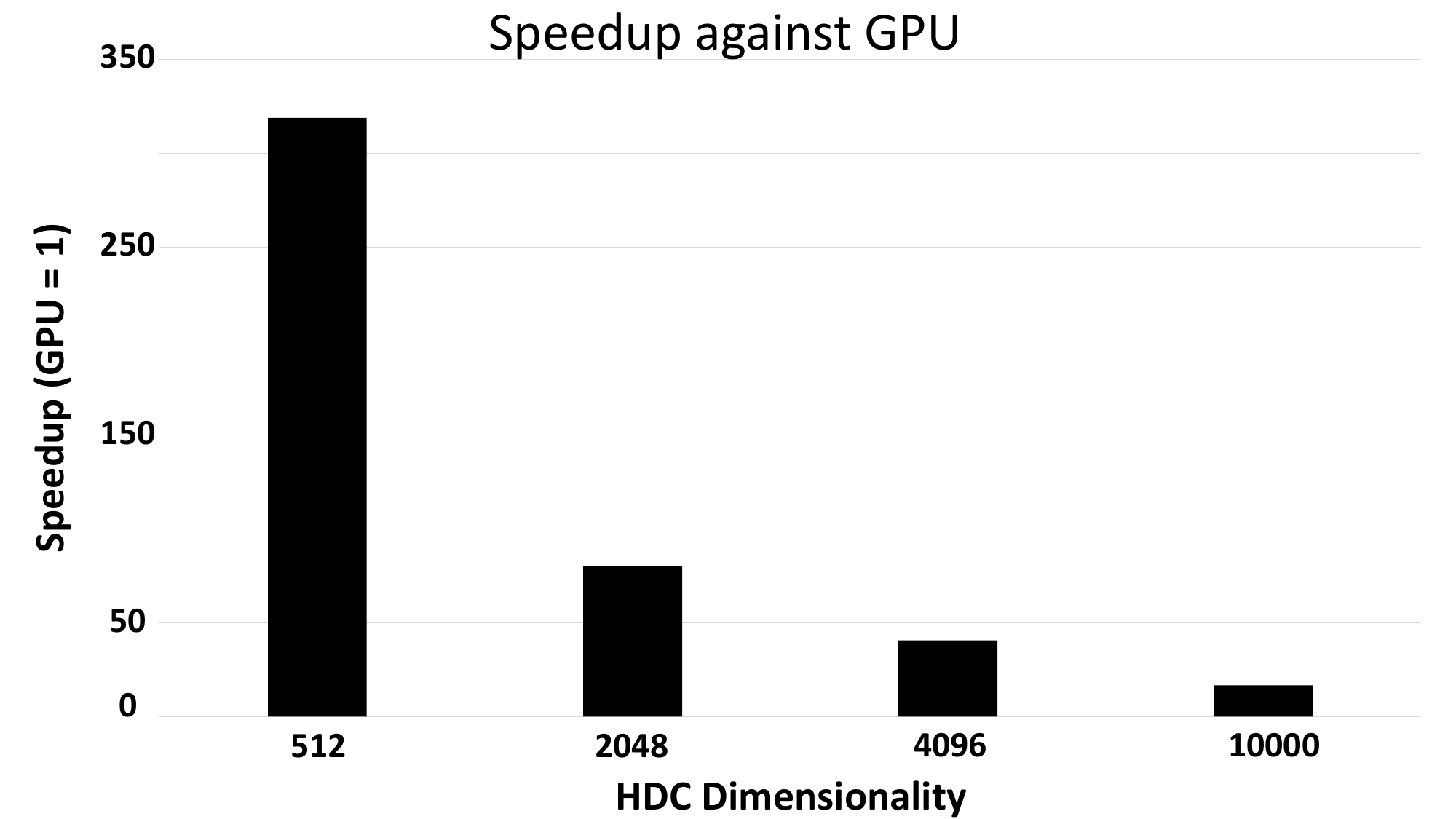}
    \caption{Speedup of our framework against NVIDIA GPU RTX 4070 for the CIFAR-1O and FMNIST datasets respectively.}
    \label{fig:speedup_dist}
\end{figure*}

\begin{figure*}[t!]
    \centering
    \includegraphics[width=0.85\textwidth]{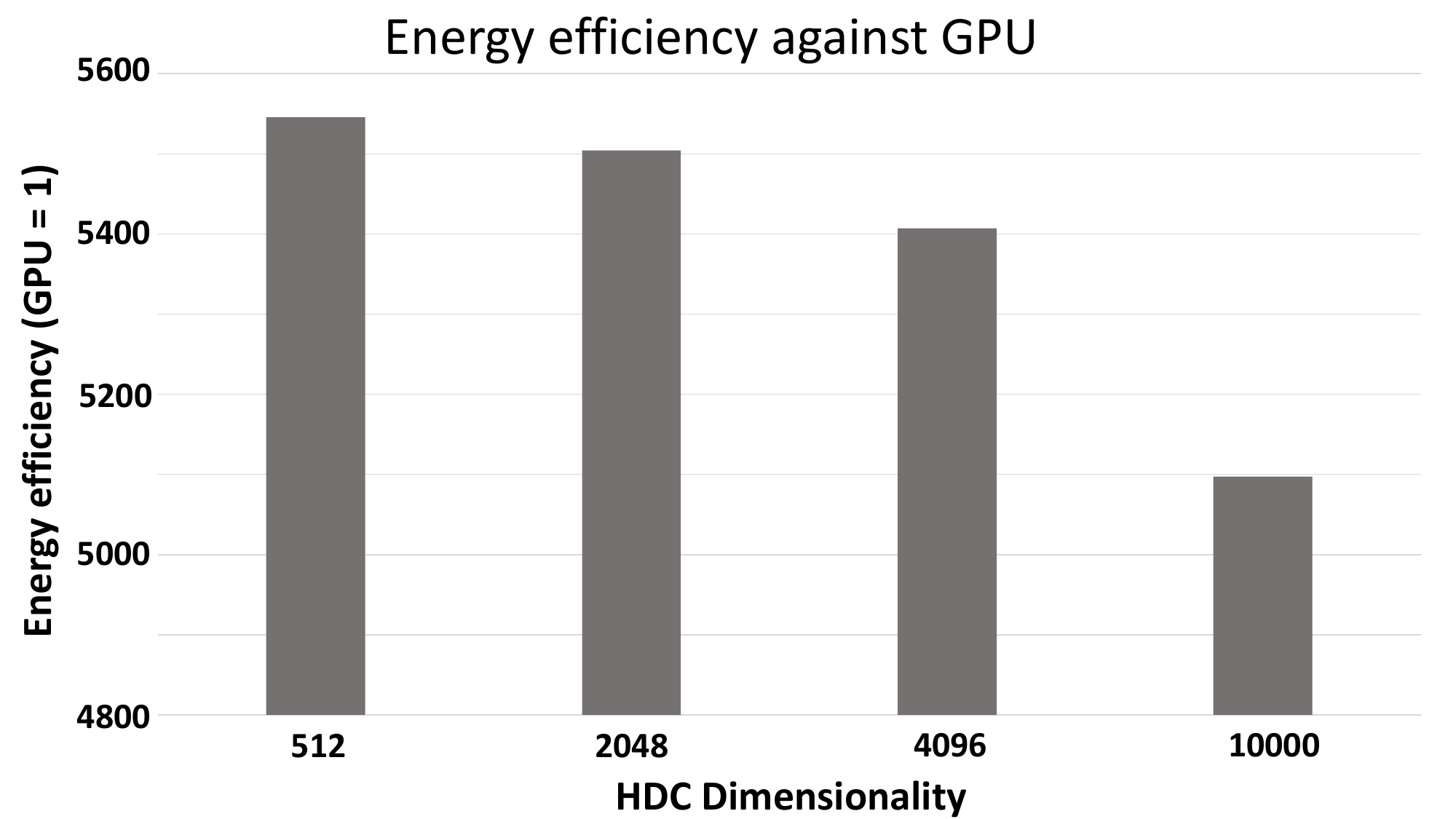}
    \caption{Energy efficiency of our framework against NVIDIA GPU RTX 4070 for the CIFAR-10 and FMNIST datasets respectively.}
    \label{fig:energy_dist}
\end{figure*}

To evaluate the efficacy of our approach, we measured the energy consumption and prediction delay for 10,000 samples in each dataset, utilizing an NVIDIA GPU RTX 4070.
A comparative analysis was conducted against the architecture based on our FeFET CiM design. 
In this evaluation, the DNN's 18 layers were parallelized, and we computed the average number of features across the layers. Our results, illustrated in Figure~\ref{fig:speedup_dist}, reveal a clear trend: increasing the dimensionality leads to a reduction in the speedup achieved by our proposed CiM design over GPU when compared to the DNN for both CIFAR-10 and CIFAR-100. 
However, it is noteworthy that we only require a dimensionality of 4096 to attain peak accuracy, achieving a remarkable speedup of x100. In terms of energy efficiency, Figure~\ref{fig:energy_dist} demonstrates a significant advantage in favor of our ferroelectric CiM approach, with a staggering x5100 improvement for CIFAR-10 and an impressive x2618 for FMNIST.

\begin{figure*}[t!]
    \centering
    \includegraphics[width=0.85\textwidth]{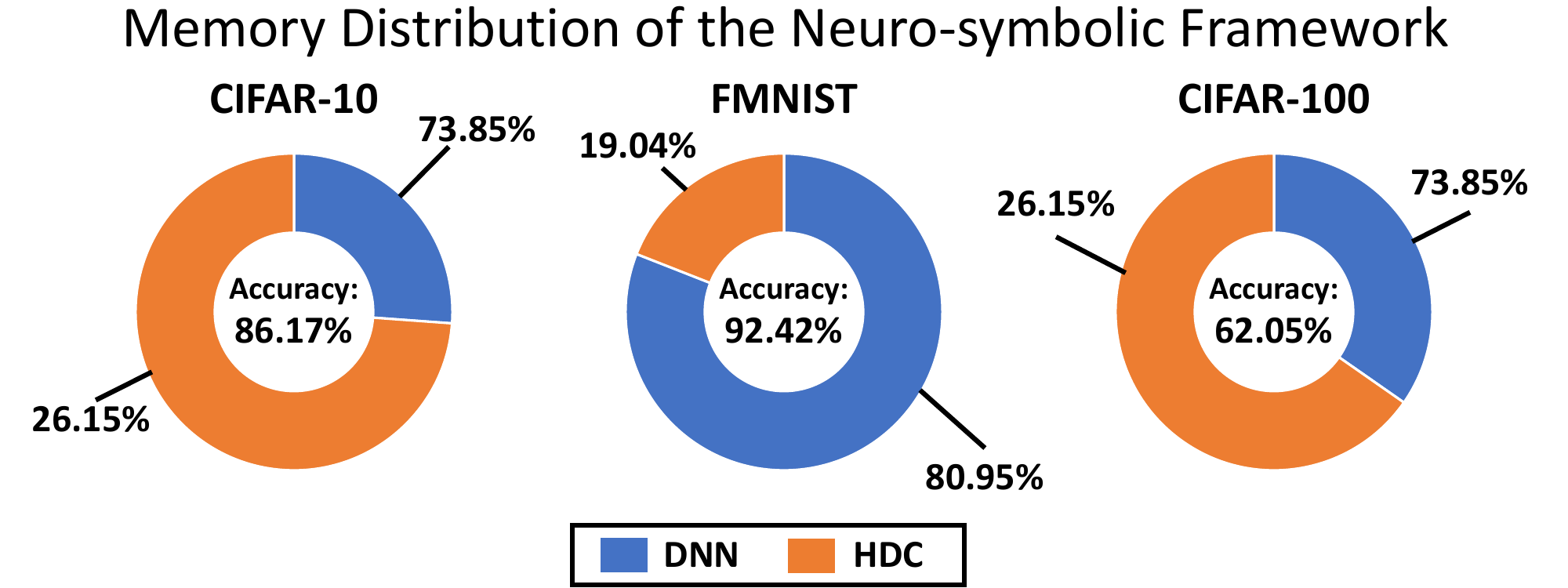}
    \caption{Memory distribution for both the DNN feature extraction model and the HDC reasoning one  to achieve the maximum accuracy, spanning 3 different datasets and tasks. The results show that for different tasks, there is a different requirement in terms of depth for the DNN layers and the dimensionality of the HDC model.}
    \label{fig:memory_dist}
\end{figure*}

Figure~\ref{fig:memory_dist} underscores the adaptability and versatility of our dual-mode CiM design across diverse tasks. The experiment demonstrates that the depth of feature extraction layers and the dimensionality of hypervectors are crucial for achieving optimal performance, and these parameters differ across tasks. 
For instance, in the case of CIFAR-10, maximum accuracy was achieved using just 9 layers and 4096 dimensionality, suggesting a heavier reliance on the reasoning component performing associative searches. 
Conversely, for FMNIST, peak accuracy was achieved with 14 layers and only 512 dimensionality, highlighting the need for superior feature extraction performing MAC operations, while the reasoning layer requires fewer resources. The CIFAR-100 dataset presents a scenario similar to CIFAR-10, with the same dimensionality but requiring an additional 5 layers to attain maximum accuracy. This experiment underscores the significance of a reconfigurable architectural design that can be easily adapted for various tasks, particularly in the context of neuro-symbolic AI.

\begin{figure*}[t!]
    \centering
    \includegraphics[width=0.85\textwidth]{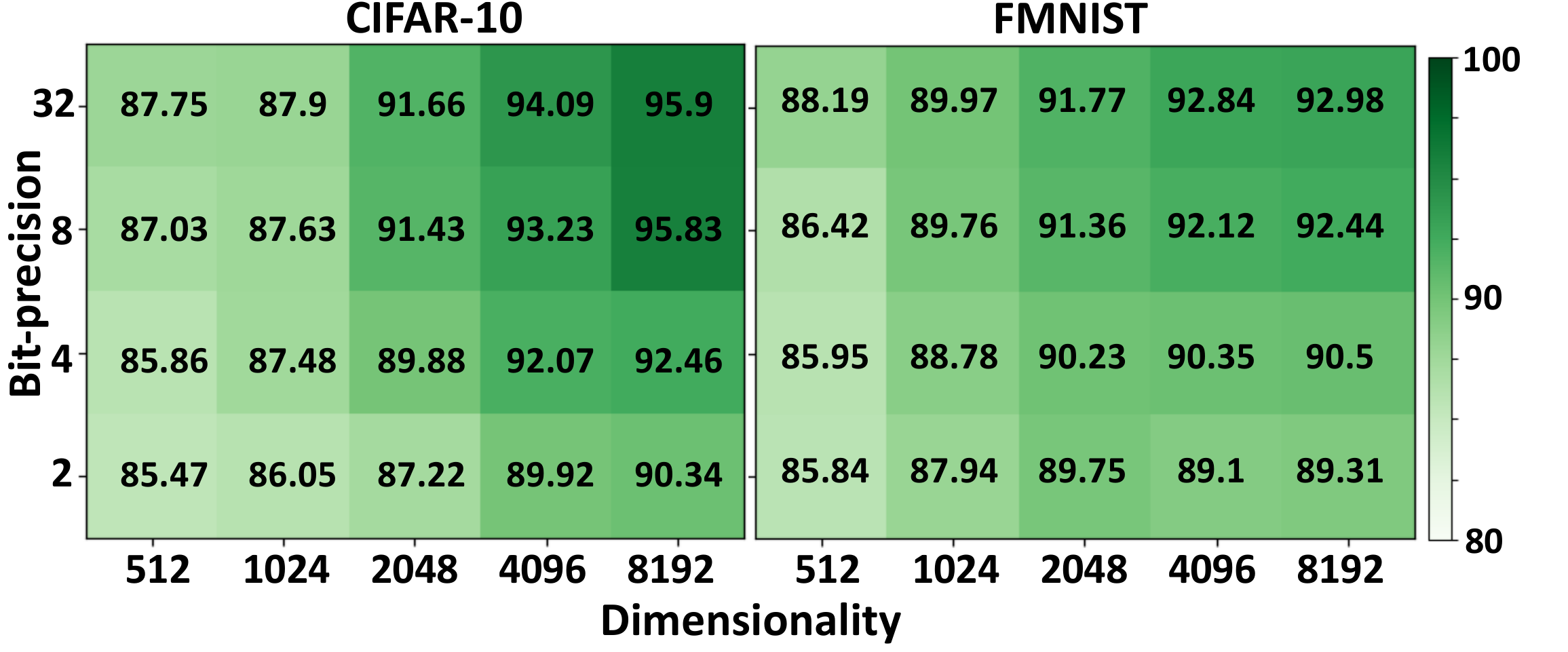}
    \caption{Accuracy of CIFAR10 and FMNIST models across different dimensions and bit-precisions under ideal conditions. The experiments vary the dimensions (512, 1024, 2048, 4096, and 8192) and bit-precisions (32, 8, 4, and 2 bits).}
    \label{fig:no_noise}
\end{figure*}

We assessed our model to understand the effects of dimensionality and bit precision after quantization using the CIFAR10 and FMNIST image datasets. 
Our experiments vary the dimensions (512, 1024, 2048, 4096, and 8192) and bit precisions (32, 8, 4, and 2 bits) under both ideal and noisy with FeFET variation conditions. 
Initially, we focused on ideal conditions, as shown in Figure~\ref{fig:no_noise}. For CIFAR10, the highest accuracy was achieved with 32-bit precision across all dimensions, peaking at 89.97\% with 1024 dimensions. 
Accuracy decreased with lower bit precisions, but the decrease was minimal compared to the substantial computational and memory savings. For instance, accuracy dropped from 87.75\% with 32 bits to 85.47\% with 2 bits at 512 dimensions, indicating a small trade-off between precision and accuracy. 
FMNIST showed similar results, with accuracy declining from 88.19\% (32 bits) to 85.84\% (2 bits) at 512 dimensions. 
This trend highlights the importance of higher bit precision in preserving model accuracy, especially for lower-dimensional representations. However, with higher dimensionalities, the impact of bit precision, such as 8 or 4 bits, was negligible, with results remaining approximately the same.

\begin{figure*}[t!]
    \centering
    \includegraphics[width=0.48\textwidth]{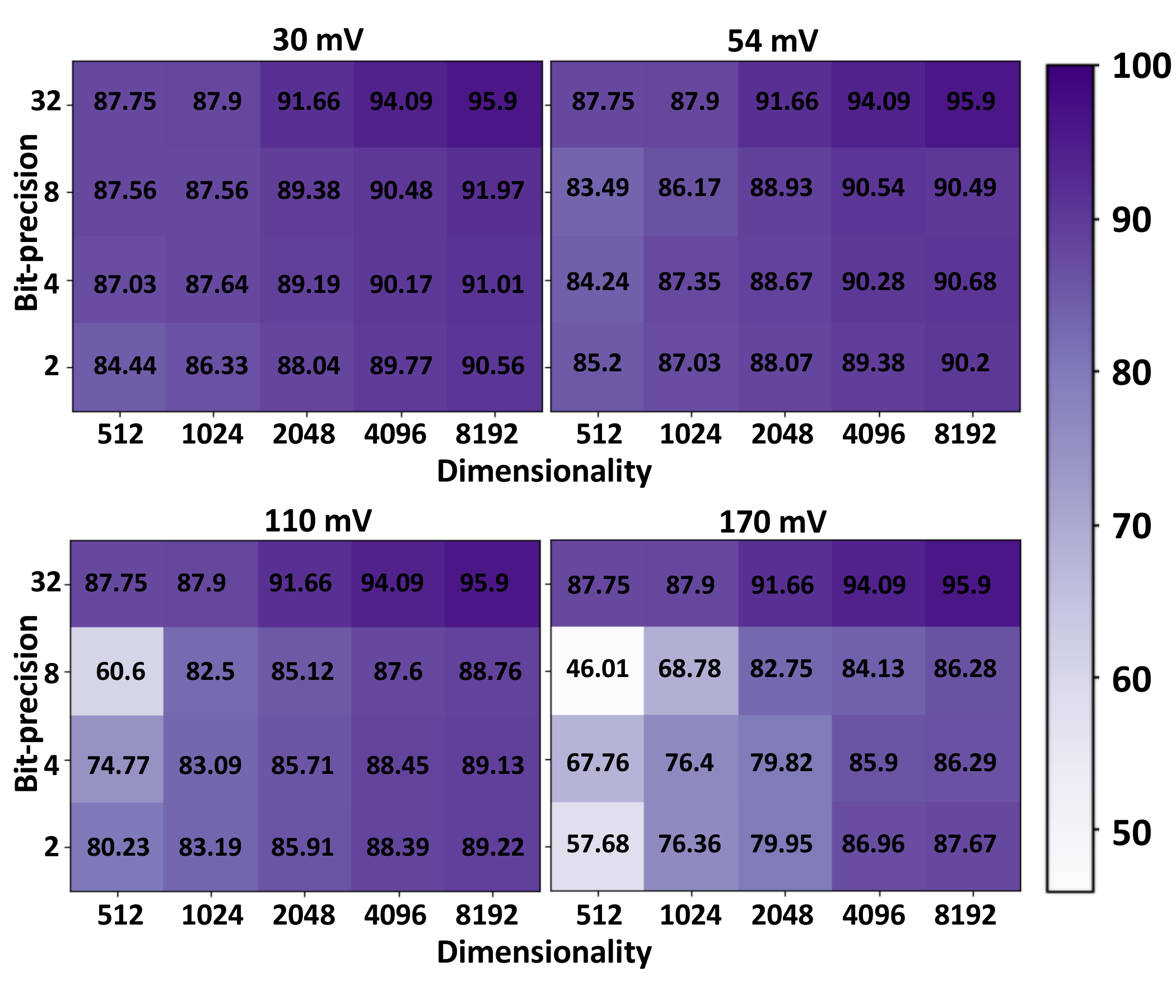}
    \includegraphics[width=0.48\textwidth]{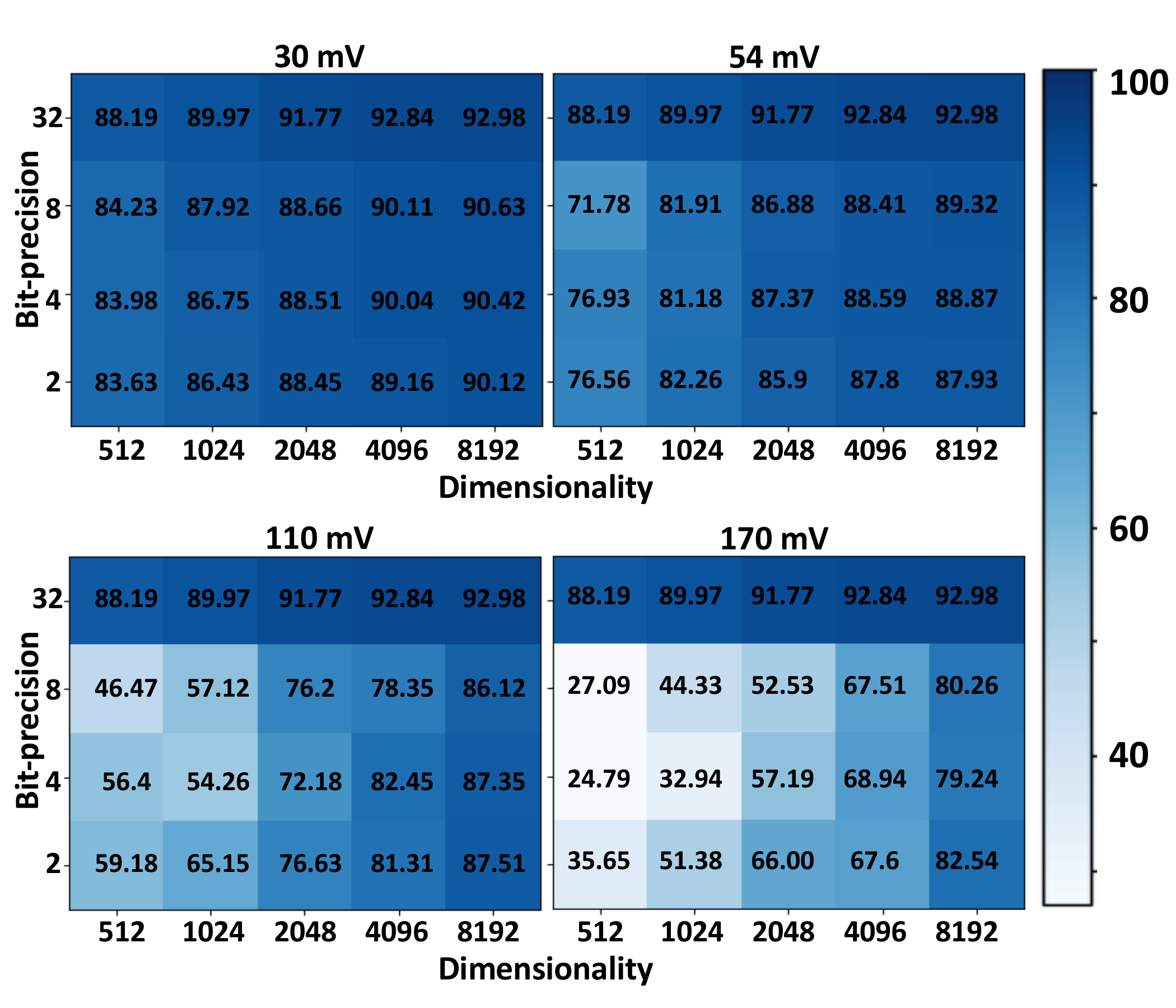}
    \caption{Impact of noise on model Accuracy across different Dimensions and bit-precisions. This heatmap illustrates the accuracy of models tested on CIFAR10 and FMNIST datasets under various noise levels (standard deviations of 0.406, 0.64045, 1.281, and 1.9). The axes represent different dimensions (512, 1024, 2048, 4096, and 8192) and bit precisions (32, 8, 4, and 2 bits). The full precision case does not consider noise as benchmark comparison.}
    \label{fig:noises}
\end{figure*}

Next, we examined the impact of device variation within hardware, as shown in Figures ~\ref{fig:noises}, starting with a low standard deviation  of FeFET threshold voltage (0.406 for 30mV case). 
For CIFAR10 at 512 dimensions, accuracy decreased from 87.75\% with 32-bit precision to 84.44\% with 2-bit precision, with quality loss due to the computation deviation caused by threshold voltage variation.
Similar patterns were observed for FMNIST, where accuracy fell from 88.19\% (32 bits) to 83.63\% (2 bits) at 512 dimensions. 
These findings indicate the detrimental effects of reduced bit precision under noisy with hardware variation conditions. 
When dimensionality was increased, the gap between the full-precision and 2-bit models widened, with quality loss up to 5.34\% for 8192 dimensions.
At a higher device variation level (0.64045 for 54 mV), CIFAR10 accuracy at 512 dimensions showed lower accuracies for higher bit precisions, a trend that persisted as the standard deviation increased. 
For example, accuracy changed from 83.49\% (8 bits) to 85.2\% (2 bits), with quality loss decreasing from 4.26\% to 2.55\% at 512 dimensions.
As dimensionality increased to 1024, quality loss decreased to 1.73\% for 8 bits and 0.87\% for 2 bits, showing a clear dependency on dimensionality rather than bit precision. 
FMNIST exhibited similar trends, with higher dimensions consistently showing better robustness to device variation compared to higher bit precisions.
With a noticeable standard deviation of 1.281 for the 110 mV case, CIFAR10 at 512 dimensions saw 8-bit precision accuracy at 60.6\%, improving to 80.23\% at 2-bit precision, with quality loss reducing from 27.15\% to 7.52\%. FMNIST at the same variation level showed accuracy improvements from 46.47\% (8 bits) to 59.18\% (2 bits), with quality loss decreasing from 41.72\% to 29.01\%.
These evaluations reveal that under high standard deviation conditions, lower bit precisions achieves better immunity to vairation at lower dimensionalities because higher bit precision suffers from small distances between symbolic vectors, leading to greater susceptibility to variation. 
However, as dimensionality increases, quality loss improves significantly, illustrating the holographic properties of HDC that effectively mitigate noise by distributing independent information in each dimension of the hypervector. 
At the highest variation level of 170 mV (std = 1.9), the same trend continued.
For CIFAR10 at 512 dimensions, 8-bit precision accuracy was 46.01\%, improving to 57.68\% at 2-bit precision, with quality loss decreasing from 41.74\% to 30.07\%. For higher dimensions such as 8192, quality loss at 8 bits improved to just 9.62\%, and 8.23\% for 2 bits. FMNIST showed similar trends. Increasing dimensionality further enhanced device variation resilience, showcasing the benefits of HDC and the effectiveness of our encoding methodology.

\end{document}